\documentclass[review]{elsarticle}

\usepackage{graphicx}
\usepackage{amsmath}
\usepackage{amsfonts}
\usepackage{array}
\usepackage{multirow}
\usepackage{lineno}
\usepackage{hyperref}
\usepackage{cleveref}
\usepackage{makecell}
\usepackage[x11names]{xcolor}

\journal{Journal of \LaTeX\ Templates}

\newif\ifproofread

\newcommand{\pfmarker}[1]{%
\ifproofread
\textcolor{red}{#1}%
\else
#1%
\fi
}

\newif\ifminorR
\newcommand{\pfminor}[1]{%
\ifminorR
\textcolor{red}{#1}%
\else
#1%
\fi
}

\newif\ifminorRNew
\newcommand{\pfminorNew}[1]{%
\ifminorRNew
\textcolor{red}{#1}%
\else
#1%
\fi
}

\begin{document}

\begin{frontmatter}

\title{Rank-1 Constrained Multichannel Wiener Filter\\ for Speech Recognition in Noisy Environments}


\author[mymainaddress]{Ziteng Wang\corref{mycorrespondingauthor}}
\cortext[mycorrespondingauthor]{Corresponding author}
\ead{wangziteng@hccl.ioa.ac.cn}

\author[mysecondaryaddress]{Emmanuel Vincent}
\author[mythirdaddress]{Romain Serizel}
\author[mymainaddress]{Yonghong Yan}

\address[mymainaddress]{University of Chinese Academy of Sciences, Beijing, 100190, China}
\address[mysecondaryaddress]{Inria, F-54600, Villers-l{\`e}s-Nancy, France}
\address[mythirdaddress]{Universit\'e de Lorraine, LORIA, UMR 7503, Vand{\oe}uvre-l\`es-Nancy, F-54506, France}

\begin{abstract}
Multichannel linear filters, such as the Multichannel Wiener Filter (MWF) and the Generalized Eigenvalue (GEV) beamformer are popular signal processing techniques which can \pfmarker{improve speech} recognition performance. In this paper, we present an experimental study on these linear filters in a specific speech recognition task, namely the CHiME-4 challenge, which features real recordings in multiple noisy environments. Specifically, the rank-1 MWF is employed for noise reduction and a new constant residual noise power constraint is derived which \pfmarker{enhances} the recognition performance. To fulfill the underlying rank-1 assumption, the speech covariance matrix is reconstructed based on eigenvectors or generalized eigenvectors. Then the rank-1 constrained MWF is evaluated with alternative multichannel linear filters under the same framework, which involves a Bidirectional Long Short-Term Memory (BLSTM) network for mask estimation. The proposed filter outperforms alternative ones, \pfminor{leading to} a 40\% relative Word Error Rate (WER) reduction compared with the baseline Weighted Delay and Sum (WDAS) beamformer on the real test set, and a 15\% relative WER reduction compared with the GEV-BAN method. The results also suggest that the speech recognition accuracy correlates more with the \pfmarker{Mel-frequency cepstral coefficients (MFCC) feature} variance than with the noise reduction or the speech distortion level.
\end{abstract}


\end{frontmatter}


\section{Introduction}

\pfmarker{Robust machine speech} recognition in real environments is a common interest for the signal processing and speech recognition communities~\cite{2014asr}. It has been a challenging task for decades. One main reason is that the target speech \pfmarker{is corrupted by} various background noises. Signal processing methods are able to extract the desired source from corrupted measurements and to improve the recognition accuracy. For this purpose, multichannel techniques improve over single-channel techniques by exploiting information not only in the time-frequency domain but also in the spatial domain. 

Multichannel linear filters, also known as beamformers, have been amply investigated in the literature~\cite{book2009benesty,book2013micarrays}. Nevertheless, only a few approaches \pfmarker{have found} widespread use in the speech recognition community until \pfmarker{recently; these include} the Weighted Delay and Sum (WDAS) beamformer in BeamformIt~\cite{2007beamformit} and the Minimum Variance Distortionless Response (MVDR) beamformer in BTK\footnote{http://distantspeechrecognition.sourceforge.net}. Recent works have explored more extensive beamforming implementations in the scope of speech recognition~\cite{2012micASR,2013reverb,2015chime3}, and the outcomes of these works indeed benefit \pfmarker{both signal processing and speech recognition communities}. On the one hand, multichannel algorithms designed to suppress noise~\cite{2001SEsharon}, reverberation~\cite{2012WPE} or competing speech, can be used as preprocessing steps for speech recognition. Though they are in general intended for improving the speech perceptual quality~\cite{2009hearaids}, some improvements are typically \pfmarker{also achieved in terms of speech recognition performance}. On the other hand, the speech recognition application inspires many new beamforming architectures~\cite{2016deepbeamforming,2016NAbf}. The recognition accuracy metric can also highlight an algorithm from a different perspective~\cite{2016icasspBLSTM}.

Remarkably, Deep Neural Network (DNN) based linear filtering has gained popularity with its success in recent speech recognition challenges~\cite{2015inriaCHiME3,2016arieCHiME,2015blstmCHiME3,2016lstmCHiME4}. A regression DNN can be used to predict the speech spectra and combined with the classical multichannel Gaussian model to derive a Multichannel Wiener Filter (MWF)~\cite{2015inriaCHiME3,2016arieCHiME}. Alternatively, a Bi-directional Long Short-Term Memory (BLSTM) network can be applied as a classification model to predict a spectral mask and combined with the MVDR beamformer or the Generalized Eigenvalue (GEV) beamformer~\cite{2015blstmCHiME3,2016lstmCHiME4}. The mask is used in the calculation of the source covariance matrix, from which the linear filter coefficients are obtained. Deep neural networks have proved to be more capable of estimating the speech second-order statistics or the speech presence probability than traditional methods.

Among the above linear filters, the MVDR beamformer is theoretically designed to be distortionless~\cite{1987MVDR}, while the GEV beamformer is targeted to achieve maximum output Signal-to-Noise Ratio (SNR)~\cite{2007GEV}. MWF~\cite{2002MWF} is a Minimum Mean Square Error (MMSE) solution which allows for given noise reduction at the expense of some speech distortion. There exist other linear filter variants, such as the Speech Distortion Weighted MWF (SDW-MWF)~\cite{2004sdwmwf,2007sdwmwf2,2014serizellow} and the Variable Span (VS) linear filter~\cite{2016vsfilter}. The SDW-MWF involves a trade-off parameter which tunes the speech distortion versus the noise reduction. In the case of a single target source, it can be expressed in the form of a spatial-prediction MWF~\cite{2008spMWF} or a rank-1 MWF~\cite{2010r1MWF}. Note that these linear filters are all equivalent up to a scaling factor if formulated in a unified framework~\cite{2016vsfilter,2013benestyNRtime,2016consolidated}. While the speech quality performance of these filters has been well studied, the comparison in terms of speech recognition performance is lacking. An interesting question is whether the already known speech quality performance can be related to the speech recognition accuracy.

\begin{figure}[t!]
	\centering
	\includegraphics[width=0.8\linewidth]{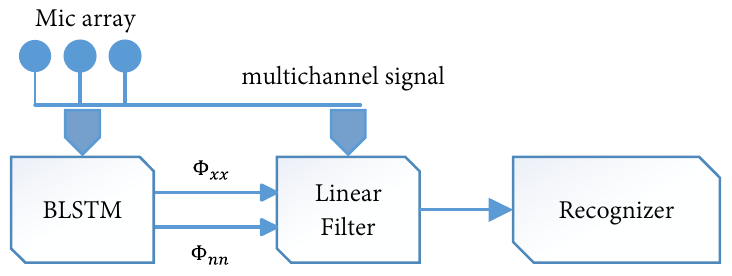}
	\caption{System illustration with BLSTM supported linear filters. ${\bf \Phi}_{xx}$ is the speech covariance matrix and ${\bf \Phi}_{nn}$ is the noise covariance matrix.}
	\label{fig1}
\end{figure}

In this paper, we provide an extensive experimental study of the relative performance of these multichannel linear filters, considering the real world speech recognition task in multiple noisy environments of the CHiME-4 challenge~\cite{2016chime4}. In particular, we focus on a family of rank-1 MWF variants. We propose a new constraint of constant residual noise power along both time and frequency, which links the rank-1 MWF and the GEV beamformer. This constraint is shown to enhance the speech recognition performance. To fulfill the underlying rank-1 assumption, we introduce a speech covariance matrix reconstruction process. The reconstruction is based on eigenvectors or generalized eigenvectors. In the experiments, all linear filters are supported by the same BLSTM network, which is used for mask estimation. An overview of the system is given in Fig.~\ref{fig1}. We also introduce a novel feature variance metric that correlates well with the Word Error Rate (WER) and helps understanding the benefit of the proposed constant residual noise power constraint.

The rest of this paper is organized as follows. The multichannel signal processing problem is formulated in Section 2. \pfmarker{In Section 3, the rank-1 MWF solution is first introduced. Three filter variants, including the novel constant residual noise power filter, are then derived separately. To fulfill the rank-1 assumption in practice, the eigenvector based speech covariance matrix reconstruction is discussed in Section 4.} The speech recognition experiments, the BLSTM network for mask estimation, the results and the analysis are presented in Section 5. Conclusions are drawn in Section 6.

\section{Problem formulation}

The multichannel signal processing problem is formulated as follows. A target speech source $s$ propagates in the acoustic space and impinges on an array of $M$ microphones. The observations at time $t$ are given by
\begin{equation}\label{eq1}
  y_m(t)=g_m*s(t)+n_m(t), \quad  m=1,2,...,M
\end{equation}
where $*$ denotes convolution, $g_m$ is the time-invariant acoustic impulse response from the source to the $m$th microphone and $n_m$ is the undesired noise at microphone $m$. Under the narrowband assumption~\cite{2016consolidated}, the above model can be written in the frequency domain as
\begin{eqnarray}\label{eq2}
\nonumber Y_m(l,k)&=&G_m(k)S(l,k)+N_m(l,k) \\
&=&X_m(l,k)+N_m(l,k), \quad  m=1,2,...,M
\end{eqnarray}
where $l$ and $k$ are respectively the frame index and the frequency index. $Y_m(l,k)$, $S(l,k)$ and $N_m(l,k)$ denote the Short-Time Fourier Transform (STFT) coefficients of $y_m(t)$, $s(t)$ and $n_m(t)$, respectively, and $G_m(k)$ is the Fourier transform of $g_m$. $X_m(l,k)=G_m(k)S(l,k)$ is the narrowband approximation of the reverberated source.

Linear filtering techniques aim to design an optimal filter ${\bf h}(l,k)=[H_1(l,k),$ $ ..., H_M(l,k)]^T$ which extracts the desired source and suppresses the other components, where subscript ${}^T$ denotes transposition. This filter is applied to the observation vector ${\bf y}(l,k)=[Y_1(l,k),~...,~Y_M(l,k)]^T$, and the filter output is
\begin{eqnarray}
\nonumber  O\pfmarker{(l,k)} &=& {\bf h}^H\pfmarker{(l,k)} {\bf y}\pfmarker{(l,k)} \\
    &=& {\bf h}^H\pfmarker{(l,k)} {\bf x}\pfmarker{(l,k)} + {\bf h}^H\pfmarker{(l,k)} {\bf n}\pfmarker{(l,k)}
\end{eqnarray}
where ${}^H$ denotes Hermitian transpose, ${\bf x}(l,k)=[X_1(l,k),~...,~X_M(l,k)]^T$ and ${\bf n}(l,k)=[N_1(l,k),~...,~N_M(l,k)]^T$.

The filter coefficients can be derived by setting certain constraints on the filtered output, for instance, to achieve MMSE with respect to an arbitrary channel of the reverberated source, say $X_1(l,k)$. This is expressed as the optimization problem:
\begin{equation}\label{eq:mse0}
  \underset{\bf h}{\min} ~ E\{|{\bf h}^H\pfmarker{(l,k)}{\bf y}\pfmarker{(l,k)} - X_1\pfmarker{(l,k)}|^2\}
\end{equation}
where $E\{ \cdot \}$ means expectation. Assuming speech and noise are uncorrelated, we can rewrite~(\ref{eq:mse0}) as
\begin{equation}\label{eq:mmse0}
  \underset{\bf h}{\min} ~ E\{|{\bf h}^H\pfmarker{(l,k)} {\bf x}\pfmarker{(l,k)} - X_1\pfmarker{(l,k)}|^2\} + E\{|{\bf h}^H\pfmarker{(l,k)}{\bf n}\pfmarker{(l,k)}|^2\}
\end{equation}
where the first term is the speech distortion and the second term is the residual noise power. A weight $\mu$ can be introduced to control the contribution of the second term:
\begin{equation}\label{eq:mmse}
  \underset{\bf h}{\min} ~ E\{|{\bf h}^H\pfmarker{(l,k)} {\bf x}\pfmarker{(l,k)} - X_1\pfmarker{(l,k)}|^2\} + \mu E\{|{\bf h}^H\pfmarker{(l,k)}{\bf n}\pfmarker{(l,k)}|^2\}.
\end{equation}
The solution of this weighted optimization problem is known as the SDW-MWF~\cite{2004sdwmwf}
\begin{equation}\label{eq:sdw-mwf}
{\bf h}_{\text{SDW-MWF}}\pfmarker{(l,k)}=({\bf \Phi}_{xx}\pfmarker{(l,k)}+\mu{\bf \Phi}_{nn}\pfmarker{(l,k)})^{-1}{\bf \Phi}_{xx}\pfmarker{(l,k)}{\bf u}_{1}
\end{equation}
where ${\bf \Phi}_{xx}(l,k)=E\{{\bf x}(l,k){\bf x}^H(l,k)\}$ is the speech covariance matrix, ${\bf \Phi}_{nn}(l,k)=E\{{\bf n}(l,k){\bf n}^H(l,k)\}$ is the noise covariance matrix and ${\bf u}_1=[1, ~0, ~...,~0]^T$ is an $M$-dimensional vector that projects on the first channel. \pfmarker{The hyperparameter $\mu$ in the SDW-MWF controls the trade-off between speech distortion and noise reduction. A larger value of $\mu$ leads to more noise reduction at the expense of more speech distortion. Specially, the plain MWF is obtained with $\mu=1$.}

\section{Rank-1 MWF variants}

\pfmarker{In the following, we first review the rank-1 MWF solution~\cite{2010r1MWF}. Then three filter variants, namely the minimum distortion filter, the plain rank-1 MWF and the new constant residual noise power filter, are derived separately by finding the proper trade-off parameter values. While the first two variants were discussed in~\cite{2010r1MWF}, the last one is obtained here by analysing the GEV beamformer~\cite{2007GEV}, a filter that maximizes the output SNR. We show that the GEV beamformer also features a constant residual noise power property over both time and frequency. The new rank-1 MWF variant is then derived following this constraint.}

\pfmarker{\subsection{Rank-1 MWF}}

Under the narrowband approximation (2), the speech covariance matrix can be decomposed as
\begin{equation}\label{eq:phix}
  {\bf \Phi}_{xx}\pfmarker{(l,k)}=\phi_{ss}\pfmarker{(l,k)}{\bf g}\pfmarker{(k)}{\bf g}^H\pfmarker{(k)}
\end{equation}
where $\phi_{ss}$ denotes the speech power spectral density and ${\bf g}(k)=[G_1(k),...,~G_M(k)]^T$ is the vector of acoustic transfer functions. This matrix is of rank-1. Thus ${\bf \Phi}_{nn}^{-1}(l,k){\bf \Phi}_{xx}(l,k)$ is also of rank-1. Its unique non-zero eigenvalue is given by
\begin{equation}\label{eq:lamda}
  \lambda\pfmarker{(l,k)}={\text{tr}} \{ {\bf \Phi}_{nn}^{-1}\pfmarker{(l,k)}{\bf \Phi}_{xx}\pfmarker{(l,k)} \}
\end{equation}
where  ${\text{tr}}\{\cdot \}$ is the trace operation. With Woodbury's identity and the fact that
\begin{equation}\label{eq:tr}
  {\bf g}^H\pfmarker{(k)}{\bf \Phi}_{nn}^{-1}\pfmarker{(l,k)}{\bf g}\pfmarker{(k)}= {\text{tr}}\{{\bf \Phi}_{nn}^{-1}\pfmarker{(l,k)}{\bf g}\pfmarker{(k)}{\bf g}^H\pfmarker{(k)} \}
\end{equation}
the SDW-MWF solution ends up in the rank-1 MWF
\begin{equation}\label{eq:r1mwf}
{\bf h}_{\text{r1MWF}-\mu}\pfmarker{(l,k)}=\frac{{\bf \Phi}_{nn}^{-1}\pfmarker{(l,k)}{\bf \Phi}_{xx}\pfmarker{(l,k)}}{\mu+\lambda\pfmarker{(l,k)}}{\bf u}_{1}.
\end{equation}
\pfmarker{Similarly, the trade-off parameter $\mu$ controls the speech distortion and noise reduction performance. With different parameter values, the corresponding filter variants exhibit different properties.}

\pfmarker{\subsection{Minimum distortion filter and plain rank-1 MWF}}

\pfmarker{These two filter variants match the cases of $\mu=0$ and $\mu=1$, respectively:}
\begin{equation}\label{eq:hdistless}
  {\bf h}_{\text{r1MWF}-0}\pfmarker{(l,k)}=\frac{{\bf \Phi}_{nn}^{-1}\pfmarker{(l,k)}{\bf \Phi}_{xx}\pfmarker{(l,k)}}{\lambda\pfmarker{(l,k)}}{\bf u}_{1},
\end{equation}
\begin{equation}\label{eq:hwiener}
  {\bf h}_{\text{r1MWF}-1}\pfmarker{(l,k)}=\frac{{\bf \Phi}_{nn}^{-1}\pfmarker{(l,k)}{\bf \Phi}_{xx}\pfmarker{(l,k)}}{1+\lambda\pfmarker{(l,k)}}{\bf u}_{1}.
\end{equation}
\pfmarker{${\bf h}_{\text{r1MWF}-0}$ is indeed distortionless in theory.}

\pfmarker{\subsection{Constant residual noise power filter}}

\pfmarker{To derive the new filter variant,} we first investigate the maximum SNR filter that is defined as
\begin{equation}\label{eq:gev-def}
{\bf h}\pfmarker{(l,k)}=\underset{\bf h}{\text{argmax}} \; \frac{{\bf h}^H\pfmarker{(l,k)}{\bf \Phi}_{xx}\pfmarker{(l,k)}{\bf h}\pfmarker{(l,k)}}{{\bf h}^H\pfmarker{(l,k)}{\bf \Phi}_{nn}\pfmarker{(l,k)}{\bf h}\pfmarker{(l,k)}}.
\end{equation}
This is a generalized Rayleigh quotient and the GEV solution is
\begin{equation}\label{eq:gev}
{\bf h}_{\text{GEV}}\pfmarker{(l,k)}= \mathcal{P} \{ {\bf \Phi}_{nn}^{-1}\pfmarker{(l,k)}{\bf \Phi}_{xx}\pfmarker{(l,k)} \}
\end{equation}
where $\mathcal{P}\{ \cdot \}$ takes the eigenvector corresponding to the largest eigenvalue, which is defined up to an arbitrary scale. An additional Blind Analytical Normalization (BAN) post-filter can be applied to control the speech distortion~\cite{2007GEV}. The output SNR of the GEV beamformer is equal to the largest eigenvalue of ${\bf \Phi}_{nn}^{-1}(l,k){\bf \Phi}_{xx}(l,k)$, which is exactly $\lambda$ in the rank-1 case.

Meanwhile, the two Hermitian matrices ${\bf \Phi}_{xx}(l,k)$ and ${\bf \Phi}_{nn}(l,k)$ can be jointly diagonalized as
\begin{equation}\label{eq:jointdiag}
\left \{ \begin{matrix}
{\bf B}^H{\bf \Phi}_{xx}\pfmarker{(l,k)}{\bf B}={\bf \Lambda}  \\
{\bf B}^H{\bf \Phi}_{nn}\pfmarker{(l,k)}{\bf B}={\bf I} ~
\end{matrix}\right.
\end{equation}
where ${\bf B}$ and ${\bf \Lambda}$ are respectively the eigenvector\footnote{Note that the eigenvectors are not of unit norm here: they are scaled such that ${\bf B}^H{\bf \Phi}_{nn}(l,k){\bf B}={\bf I}$ holds.} and eigenvalue matrices of ${\bf \Phi}_{nn}^{-1}(l,k){\bf \Phi}_{xx}(l,k)$, and {\bf I} is the identity matrix~\cite{2016vsfilter}. If the diagonal elements of ${\bf \Lambda}$ are in descending order, then the GEV beamformer~(\ref{eq:gev}) can be chosen as the first column vector of ${\bf B}$. This is the usual choice made in the literature~\cite{2007GEV} and the one we \pfmarker{also make in the following}. We \pfmarker{denote it by} ${\bf h}^{*}_{\text{GEV}}(l,k)= \mathcal{P}^{*} \{ {\bf \Phi}_{nn}^{-1}(l,k){\bf \Phi}_{xx}(l,k) \}$. By defining the residual noise as $\xi_n={\bf h}^H{\bf n}$, we see that the residual noise power of the GEV is given by
\begin{equation}\label{eq:rnp1}
  E\{|\xi_n\pfmarker{(l,k)}|^2\}={{\bf h}^{*}}^H_{\text{GEV}}\pfmarker{(l,k)}{\bf \Phi}_{nn}\pfmarker{(l,k)}{\bf h}^{*}_{\text{GEV}}\pfmarker{(l,k)}=1,
\end{equation}
which indicates constant residual noise power over both frequency and time.

Going back to the rank-1 MWF, \pfmarker{it can be proved that the rank-1 MWF solution also satisfies~(\ref{eq:gev-def}) with an arbitrary trade-off parameter}. The general expectation of the residual noise power is
\begin{displaymath}
E\{|\xi_n\pfmarker{(l,k)}|^2\}={\bf h}^H_{\text{r1MWF}}\pfmarker{(l,k)}{\bf \Phi}_{nn}\pfmarker{(l,k)}{\bf h}_{\text{r1MWF}}\pfmarker{(l,k)} \qquad
\end{displaymath}
\begin{eqnarray}\label{eq:rnp}
\nonumber &=& \frac{{\bf u}_{1}^T{\bf \Phi}_{xx}\pfmarker{(l,k)}{\bf \Phi}_{nn}^{-1}\pfmarker{(l,k)}{\bf \Phi}_{xx}\pfmarker{(l,k)}{\bf u}_{1}}{(\mu+\lambda\pfmarker{(l,k)})^2} \\
\nonumber   &=& \frac{\phi_{x_1x_1}\pfmarker{(l,k)}\phi_{ss}\pfmarker{(l,k)}{\bf g}^H\pfmarker{(k)}{\bf \Phi}_{nn}^{-1}\pfmarker{(l,k)}{\bf g}\pfmarker{(k)}}{(\mu+\lambda\pfmarker{(l,k)})^2} \\
   &=& \frac{\phi_{x_1x_1}\pfmarker{(l,k)}\lambda\pfmarker{(l,k)}}{(\mu+\lambda\pfmarker{(l,k)})^2}
\end{eqnarray}
in which the final step makes use of equation (\ref{eq:tr}). Setting the residual noise power to a constant value $E\{|\xi_n(l,k)|^2\} = 1$ as in~(\ref{eq:rnp1}), and taking it into equation (\ref{eq:rnp}), we obtain
\begin{equation}
  \mu_{\text G}\pfmarker{(l,k)} = \sqrt{\phi_{x_1x_1}\pfmarker{(l,k)}\lambda\pfmarker{(l,k)}} - \lambda\pfmarker{(l,k)}
\end{equation}
\pfminor{which has become frame and frequency dependent.} Thus a rank-1 MWF filter which is similar to the GEV in terms of maximizing the output SNR and leading to constant residual noise power, but different in terms of projection direction, is given by
\begin{equation}\label{eq:hgev}
  {\bf h}_{\text{r1MWF}-\mu_{\text{G}}}\pfmarker{(l,k)}=\frac{{\bf \Phi}_{nn}^{-1}\pfmarker{(l,k)}{\bf \Phi}_{xx}\pfmarker{(l,k)}}{\mu_{\text G}\pfmarker{(l,k)} + \lambda\pfmarker{(l,k)}}{\bf u}_{1}.
\end{equation}
This choice of $\mu=\mu_{\text G}$ is new in the context of rank-1 MWF. \pfmarker{Although it has been known that linear filters are equivalent up to a scaling factor~\cite{2016vsfilter,2013benestyNRtime,2016consolidated}, the factor that specifically relates the rank-1 MWF and GEV is given here by $\frac{1}{\mu_{\text G} + \lambda}$ for the first time.}

In~\cite{2015residual}, the residual noise power was chosen as constant over time. Here we restrict it to be constant along frequency too. Note that, under this constraint, the signal can be amplified in some noise-dominated frequency bins and weakened in some speech-dominated frequency bins, which induces speech distortion \pfminor{as does the} GEV beamformer. Nevertheless, the derived three rank-1 MWF variants differ only by the spectral shape of the filtered signal. They all project in the spatial direction of ${\bf \Phi}_{nn}^{-1}(l,k){\bf \Phi}_{xx}(l,k){\bf u}_{1}$, \pfmarker{but with different spectral gains}.

\pfmarker{\section{Rank-1 constraint on the speech covariance matrix}}

The above linear filters are specified as functions of the covariance matrices: ${\bf \Phi}_{xx}(l,k)$ and ${\bf \Phi}_{nn}(l,k)$. In practice, the covariance matrices need to be estimated either by recursive smoothing
\begin{eqnarray}\label{eq:recsmoo}
  {\bf \tilde\Phi}_{xx}(l,k) &=& \alpha{\bf \tilde\Phi}_{xx}(l-1,k) + (1-\alpha)\pfmarker{\mathcal{M}_{x}(l,k){\bf y}(l,k){\bf y}^H(l,k)} \\
  {\bf \tilde\Phi}_{nn}(l,k) &=& \alpha{\bf \tilde\Phi}_{nn}(l-1,k) + (1-\alpha)\pfmarker{\mathcal{M}_{n}(l,k){\bf y}(l,k){\bf y}^H(l,k)}
\end{eqnarray}
or by the arithmetic mean
\begin{eqnarray}\label{eq:armean}
  {\bf \tilde\Phi}_{xx}(l,k) &=& \frac{1}{L}\sum_{l=-L/2}^{L/2-1}\pfmarker{\mathcal{M}_{x}(l,k){\bf y}(l,k){\bf y}^H(l,k)} \\
  {\bf \tilde\Phi}_{nn}(l,k) &=& \frac{1}{L}\sum_{l=-L/2}^{L/2-1}\pfmarker{\mathcal{M}_{n}(l,k){\bf y}(l,k){\bf y}^H(l,k)}
\end{eqnarray}
where $\alpha$ is a forgetting factor, and $\mathcal{M}_{x}$, $\mathcal{M}_{n}$ represent the speech and noise masks or the speech and noise presence probabilities, respectively. Due to estimation errors or to the fact that the narrowband assumption (\ref{eq:phix}) doesn't hold perfectly, the estimated speech covariance ${\bf \Phi}_{xx}(l,k)$ is not rank-1. \pfmarker{In~\cite{2014serizellow}, using a low-rank approximation of the speech covariance matrix in the SDW-MWF effectively delivered better noise reduction performance. This motivates us to constrain the estimated speech covariance matrix to be rank-1 as follows.}

The matrix can be decomposed into a rank-1 part and a remainder part:
\begin{eqnarray}\label{eq:r1appr}
\nonumber {\bf \tilde\Phi}_{xx}\pfmarker{(l,k)} &=& {\bf \Phi}_{r1}\pfmarker{(l,k)} + {\bf \Phi}_{z}\pfmarker{(l,k)} \\
         &=& \sigma_x\pfmarker{(l,k)} {\bf a}\pfmarker{(l,k)}{\bf a}^H\pfmarker{(l,k)} + {\bf \Phi}_{z}\pfmarker{(l,k)}
\end{eqnarray}
where $\sigma_x={\text{tr}}\{{\bf \tilde\Phi}_{xx}(l,k)\}/{\text{tr}}\{{\bf a}(l,k){\bf a}^H(l,k)\}$, and ${\bf a}(l,k)$ is defined as the reconstruction vector. The remainder matrix ${\bf \Phi}_{z}(l,k)$ can be either treated as noise or simply ignored, leading to different interpretations of the filter~\cite{2014serizellow}. We choose to ignore the remainder part here. ${\bf a}(l,k)$ is chosen from the eigenvector and the generalized eigenvector, that are defined as:
\begin{eqnarray}
{\bf a}_{\text{EVD}}\pfmarker{(l,k)} &=& \mathcal{P} \{ {\bf \tilde\Phi}_{xx}\pfmarker{(l,k)} \}   \\
{\bf a}_{\text{GEVD}}\pfmarker{(l,k)} &=& \pfminorNew{{\bf \tilde\Phi}_{nn}}\mathcal{P} \{ {\bf \tilde\Phi}_{nn}^{-1}\pfmarker{(l,k)}{\bf \tilde\Phi}_{xx}\pfmarker{(l,k)} \}.
\end{eqnarray}
\pfminorNew{Note that ${\bf a}_{\text{GEVD}}$ is interpreted as the desired source relative transfer function in~\cite{2009MultichEigen}.} \pfmarker{These two expressions result in new EVD and GEVD based filters, respectively, that fulfill the rank-1 assumption used for deriving the rank-1 MWF.} The new filters are given by
\begin{equation}\label{eq:r1cmwf}
{\bf \tilde h}_{\text{r1MWF}-\mu-\text{evd/gevd}}\pfmarker{(l,k)}=\frac{{\bf \tilde\Phi}_{nn}^{-1}\pfmarker{(l,k)}{\bf \Phi}_{r1}\pfmarker{(l,k)}}{\mu+\lambda\pfmarker{(l,k)}}{\bf u}_{1}.
\end{equation}

\section{Experiments and analysis}

\subsection{The recognition task}

The experiments are conducted on the CHiME-4 challenge data~\cite{2016chime4}. This dataset features real recordings in four daily noise environments: bus, cafeteria, street junction and pedestrian area. Sentences from the Wall Street Journal (WSJ0) 5k corpus are read from a tablet device. Then the audio signals are captured by a 6-channel microphone array embedded in the tablet frame. For subsequent processing, the signals are downsampled to 16kHz. Besides the real recordings, there are also artificially generated sentences. Clean WSJ0 samples are mixed with the environment noises at similar SNRs as the real data. The whole dataset is divided into disjoint training, development and evaluation sets. In the training set, there are 1600 real and 7138 simulated sentences, about 20 hours in total. In the development set and the test set, there are 1640 and 1320 sentences for each kind of data.

The recognition system is the official challenge baseline built with the Kaldi toolkit\footnote{https://github.com/kaldi-asr/kaldi/tree/master/egs/chime4}. \pfmarker{The inputs to the DNN acoustic model are Mel-frequency Cepstral Coefficient (MFCC) features processed by feature space Maximum Likelihood Linear Regression (fMLLR) transformation.} The outputs are 1979 Hidden Markov Model (HMM) probability states. The acoustic model has 7 layers that are trained under the state level Minimum Bayes Risk (sMBR) criterion. In the decoding phase, a 3-gram Language Model (LM) is used. Recurrent neural network (RNN) LM rescoring is not applied in our experiments: this is the only difference with respect to the official baseline. The results obtained here are not meant to be compared to the best CHiME-4 results, where advanced acoustic and RNN language models are applied.

\subsection{Evaluation setup}

\begin{table}[!h]
\caption{Linear filters involved in the evaluation. They are organized in terms of the projection direction and spectral gain in order to highlight their differences or similarities. The filter $\bf h$ is given by the product of the projection direction and the spectral gain. Note that ${\bf \Phi}_{xx}$, ${\bf \Phi}_{nn}$, $\bf a$, $\sigma_x$, $\lambda$ and $\mu_{\text G}$ depend on time and frequency.}
\label{table1}
\begin{center}
\begin{tabular}{|l|c|c|c|}
  \hline
  linear filter & reference & projection direction & spectral gain \\ \hline
  MVDR          & \cite{1987MVDR}	& \multirow{2}{*}{ ${\bf \Phi}_{nn}^{-1}{\bf a}$, ${\bf a}=\mathcal{P}\{{\bf \Phi}_{xx}\}$ } & $\frac{\sqrt{{\bf a}^H{\bf a}}}{{\bf a}^H{\bf \Phi}_{nn}^{-1}{\bf a}}$\\ \cline{1-2} \cline{4-4}
  r1MWF-$\mu$-evd     &(\ref{eq:r1appr})(\ref{eq:r1cmwf}) & 	&  $\frac{\sigma_x {\bf a}^H{\bf u}_{1}}{{\mu+\lambda}}$, $\mu=1,\mu_{\text G}$  \\ \hline

  r1MWF-$\mu$   & (\ref{eq:r1mwf})	& ${\bf \Phi}_{nn}^{-1}{\bf \Phi}_{xx}{\bf u}_{1}$  & $\frac{1}{{\mu+\lambda}}$, $\mu=0,1,5,10,\mu_{\text G}$\\ \hline

  r1MWF-$\mu$-gevd    &(\ref{eq:r1appr})(\ref{eq:r1cmwf}) & \multirow{3}{*}{\makecell{ $\mathcal{P}^{*} \{ {\bf \Phi}_{nn}^{-1}{\bf \Phi}_{xx} \}$, \\ \pfminorNew{${\bf a}={\bf \Phi}_{nn}\mathcal{P}^{*} \{ {\bf \Phi}_{nn}^{-1}{\bf \Phi}_{xx} \}$} } } &  $\frac{\sigma_x {\bf a}^H{\bf u}_{1}}{{\mu+\lambda}}$, $\mu=1,\mu_{\text G}$  \\ \cline{1-2} \cline{4-4}

  GEV-BAN       & \multirow{2}{*}{(\ref{eq:gev})} &	 & BAN~[21]\\ \cline{1-1} \cline{4-4}
  GEV           &  &	 & 1 \\ \hline

  MWF       & (\ref{eq:sdw-mwf})	& $({\bf \Phi}_{xx}+{\bf \Phi}_{nn})^{-1}{\bf \Phi}_{xx}{\bf u}_{1}$ & 1   \\ \hline

  VS            & \cite{2016vsfilter} &   ${\bf a}{\bf a}^H{\bf \Phi}_{xx}{\bf u}_{1}$, ${\bf a}=\mathcal{P}^{*} \{ {\bf \Phi}_{nn}^{-1}{\bf \Phi}_{xx} \}$ & $\frac{1}{{\mu+\lambda}}$, $\mu=1$ \\  \hline
\end{tabular}
\end{center}
\end{table}

The WDAS beamformer~\cite{2007beamformit} is provided as the official baseline for CHiME-4. The linear filters involved in the evaluation are listed in Table~\ref{table1}. They are organized in terms of the projection direction and the spectral gain. GEV-BAN was the method used in the best CHiME-4 submissions~\cite{2016chime4}.

\begin{figure}[t]
	\centering
	\includegraphics[width=0.62\linewidth]{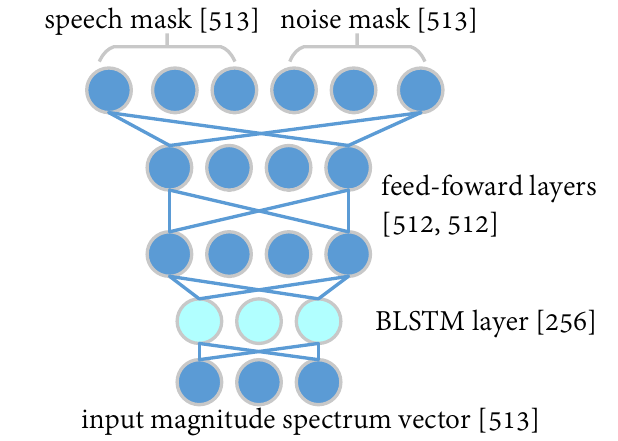}
	\caption{Illustration of the BLSTM network for mask prediction. \pfmarker{The numbers in brackets indicate the number of nodes per layer.}}
	\label{fig2}
\end{figure}

The linear filters are based on the same BLSTM network which simultaneously predicts the speech mask $\mathcal{M}_{x}$ and the noise mask $\mathcal{M}_{n}$. In~\cite{2016icasspBLSTM}, the network was combined with MVDR and GEV. We extend the process here to other linear filters. The STFT is performed in 1024 points with 256 points shift. The magnitude spectrum vector of one frame is used as input. The network consists of one recurrent BLSTM layer with 256 nodes and two feed-forward hidden layers with 512 nodes each. The outputs are 1026 nodes for the speech mask and the noise mask. The target ideal masks are defined as
\begin{equation}\label{eq:ibmx}
  \mathcal{M}_{x}=\left\{\begin{matrix}
    1 &{\rm SNR}>LC_x,  \\
    0 &{\text{otherwise}},
\end{matrix}\right.
\end{equation}
\begin{equation}\label{eq:ibmn}
\mathcal{M}_{n}=\left\{\begin{matrix}
    0 &{\rm SNR}>LC_n,  \\
    1 &{\text{otherwise}},
\end{matrix}\right.
\end{equation}
where the thresholds for speech and noise detection $LC_x$ and $LC_n$ are set to be 0 dB and -10 dB, respectively. \pfmarker{The thresholds are chosen to favor a speech/noise decision with low false acceptance rate. This results in more reliable covariance matrix estimation at the cost of discarding some time-frequency bins~\cite{2015blstmCHiME3}.} The ReLU activation function is used for all the hidden layers while the sigmoid function is chosen for the output layer. \pfmarker{The network is totally single-channel based, i.e., it operates on each microphone signal independently.} An illustration of the network architecture is shown in Fig.~\ref{fig2}.

In the training stage, the network is trained with \pfmarker{all the simulated training utterances from the 6 channels}. The simulated data from the development set is used for cross validation and early stopping. The weights of the BLSTM layer are initialized from a uniform distribution ranging from -0.05 to 0.05. The other layers are \pfmarker{initialized with samples from} a normal distribution with zero mean and a variance of $\sqrt{1/u_{\text{in}}}$ with $u_{\text{in}}$ denoting the number of input units. The Adam method~\cite{kingma2014adam} is employed to tune the network and the learning rate is adjusted adaptively. Cross-entropy loss is used as the optimization criterion. For better generalization performance, dropout is applied to all the hidden layers. The dropout rate is fixed to 0.5. Batch normalization~\cite{ioffe2015batch} is applied to speed up the training process and help the network converge to a better local optimum.

In the test phase, \pfmarker{the magnitude spectrum vector of the test signal} is fed to the trained model and the output masks are in the [0, 1] range. \pfmarker{The masks are obtained separately for each channel, and the median value is taken across channels. The median operation is robust to outliers in the case of microphone failure in the real recordings~\cite{2016icasspBLSTM}.} This value is then used to obtain ${\bf \tilde\Phi}_{xx}, {\bf \tilde\Phi}_{nn}$ using \pfminor{(\ref{eq:armean}) and (24)}. The statistics are averaged on the whole sentence, which leads to time-invariant filters \pfmarker{per utterance, that have shown to be more advantageous than time-varying ones for this speech recognition task~\cite{2016chime4}}. For the rank-1 MWF, the reference channel is decided by cross-channel correlations. The channel which has the highest average correlation score with the other channels is selected as the reference.

\pfminor{The experimental setup follows the CHiME-4 challenge instructions: no extra information, such as the environment label, is exploited.} \pfmarker{The source code is available at \url{https://github.com/ZitengWang/nn\_mask}}.

\subsection{Recognition results - Acoustic model trained on noisy data}

\begin{table*}[!th]
\caption{WERs (\%) achieved by the DNN-sMBR system trained on noisy data. The best result for each dataset is in bold.}
\label{table2}
\begin{center}
\begin{tabular}{|l|c|c|c|c||c|c|c|c|}
  \hline
Acoustic model  & \multicolumn{4}{c||}{\pfmarker{data from channel 5}} & \multicolumn{4}{c|}{\pfmarker{ data from all 6 channels}}\\ \hline
\multirow{2}{*}{Dataset}  & \multicolumn{2}{c|}{dev} & \multicolumn{2}{c||}{test} & \multicolumn{2}{c|}{dev} & \multicolumn{2}{c|}{test} \\ \cline{2-9}
       & simu & real      & simu  & real       & simu & real       & simu  & real \\ \hline \hline
Noisy channel 5  & 11.43& 12.53  & 14.15 &23.52   & 9.92& 11.00 &	11.44 & 18.86	\\ \hline
WDAS  &	~9.07 &	8.14	&	14.20 &	15.04	&	~8.09 &	7.30	&	11.97 &	12.86	\\ \hline
MVDR	       &	~6.97 &	6.86	&	~8.70 &	10.31	&	~6.21 &	6.07	&	~7.47 &	~8.89	\\ \hline
GEV-BAN	       &	~7.27 &	6.85	&	~9.17 &	10.48	&	~6.24 &	6.57	&	~8.25 &	~9.11	\\ \hline
GEV	           &	~7.54 &	7.05	&	10.01 &	10.53	&	~6.85 &	6.72	&	~9.21 &	~9.14	\\ \hline
MWF	           &	11.24 &	9.38	&	12.54 &	16.16	&	~9.48 &	7.82	&	10.17 &	13.63	\\ \hline
VS	           &	~5.41 &	6.53   &\bf ~6.37 &10.22   &\bf~4.85 &	5.58	& ~5.30 &~8.56	\\ \hline \hline

r1MWF-0	       &	~5.83 &	6.68	&	~7.03 &	11.40	&	~5.18 &	5.83	&	~5.79 &	~9.54	\\ \hline
r1MWF-1	       &	~5.86 &	6.70	&	~7.07 &	11.44	&	~5.22 &	5.84	&	~5.85 &	~9.74	\\ \hline
r1MWF-5	       &	~6.01 &	6.83	&	~7.12 &	11.71	&	~5.31 &	6.04	&	~6.00 &	10.15	\\ \hline
r1MWF-10	   &	~6.20 &	6.97	&	~7.41 &	12.00	&	~5.44 &	6.15	&	~6.21 &	10.49	\\ \hline
r1MWF-$\mu_{\text G}$ &	~6.42 &	6.43	&	~8.00 &	10.33	&	~5.76 &	5.73	&	~6.61 &	~8.89	\\ \hline \hline

r1MWF-1-evd	   &	~5.82 &	6.83	&	~7.05 &	11.17	&	~5.09 &	5.66	&	~5.99 &	~9.56	\\ \hline
r1MWF-1-gevd	&\bf ~5.37 & 6.59  & ~6.40 &	10.26  & ~4.86 &\bf 5.52	&\bf ~5.16 & ~8.47	\\ \hline
r1MWF-$\mu_{\text G}$-evd	&	~6.05 &	6.12 &	~7.63 & ~9.25 &	~5.41 &	5.54	&	~6.14 & ~8.09	\\ \hline
r1MWF-$\mu_{\text G}$-gevd	&	~6.01 &\bf 6.03	&	~6.84 &\bf ~8.74 &	~5.29 & 5.53	&	~5.83 &\bf ~7.71	\\ \hline

\end{tabular}
\end{center}
\end{table*}

In the first experiment, two acoustic models are trained with the noisy data: one with \pfmarker{utterances from the official channel 5 ($\sim$20h) and the other with utterances from all 6 channels ($\sim$120h)}. The involved linear filters are only applied to the development data and the test data. The WER results are given in Table~\ref{table2}.

From an overall perspective, the results of the linear filters follow the same trends for both acoustic models. \pfmarker{The filers consistently enhance the recognition performance} and lower WERs are achieved as expected with more training data. The performance difference between simulated data and real data is small on the development set. \pfminor{The overall higher error rates on the test set are due to the fact that the speakers of the test set speak in a less intelligible way}~\cite{2016third}. The following discussions concentrate on the results achieved on the test set with the acoustic model trained on \pfmarker{ utterances from all 6 channels}.

\begin{figure}[th!]
	\centering
	\includegraphics[width=\linewidth]{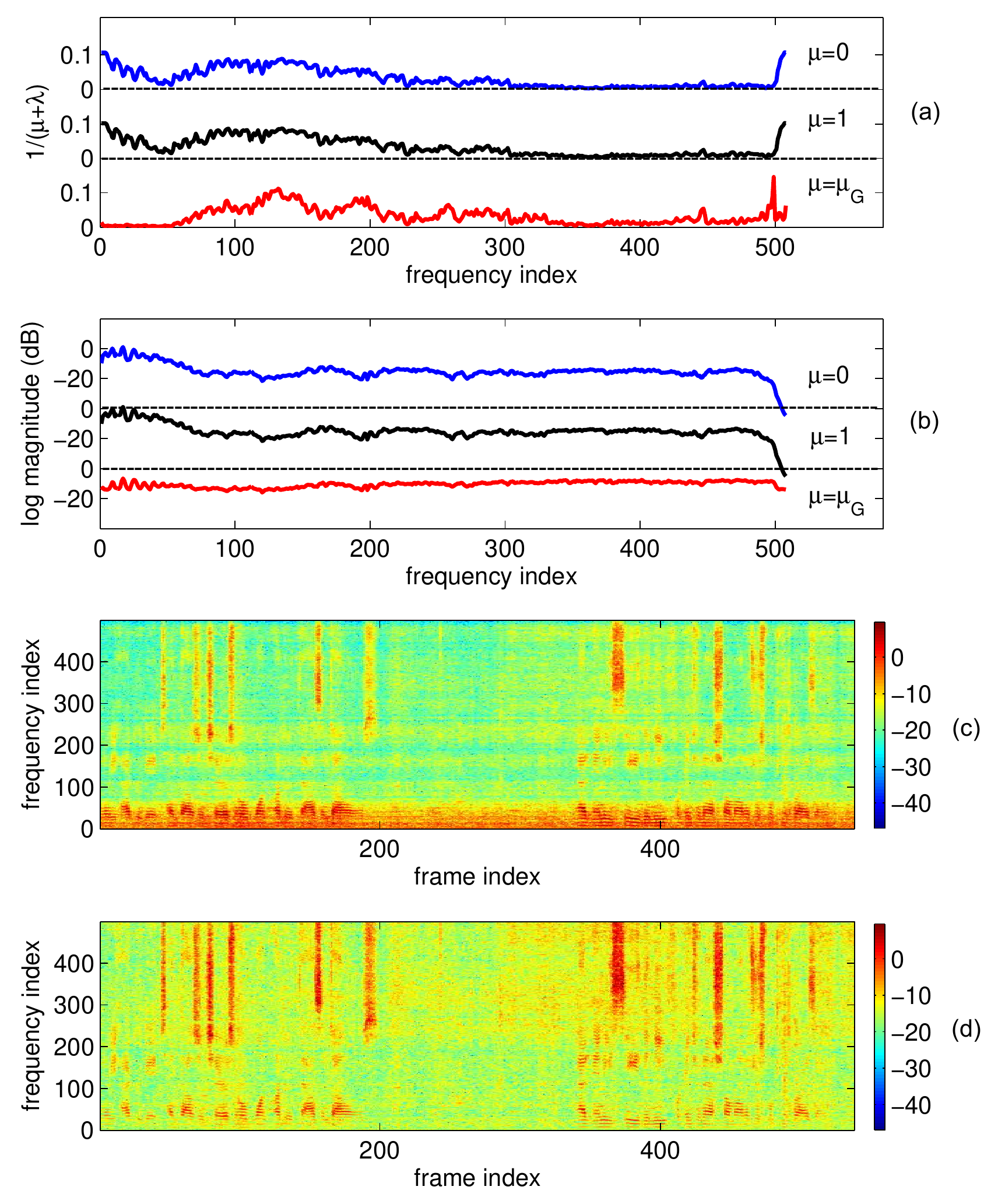}
	\caption{\pfmarker{Illustration of the r1MWF-$\mu$ filter for an example sentence (M06\_440C0201\_BUS) from the real test set.} (a) Spectral gain along frequency for different values of $\mu$. (b) The corresponding log-magnitude of one frame of the filtered signals. (c) and (d) Log-magnitude spectrograms of the filtered signals with $\mu=1$ and $\mu=\mu_G$, respectively. }
	\label{fig-mu}
\end{figure}

Compared with the noisy baseline, it is obvious that all the multichannel methods improve the speech recognition performance. The WDAS beamformer is a simple but effective technique, which delivers 35\% relative WER reduction on the real data. The MWF achieves less reduction here partly due to its sensitivity to mask estimation errors~\cite{2011performance}. The MVDR filter is theoretically speech distortionless and further improvement is achieved from the WDAS filter. For instance, the WER is reduced from 12.86\% to 8.89\% on the real data. The GEV and GEV-BAN surprisingly lead to comparable results, despite the fact that BAN is believed to be crucial to the speech perceptual quality~\cite{2007GEV}. There is around 1\% absolute difference on the simulated data though. The VS filter gets the lowest WER among the above ones. It is especially effective on the simulated data with an average 25\% relative improvement from the MVDR filter. The recognition performance is clearly influenced by the projection direction of the beamformers as shown by the GEV, MWF filters and the VS, r1MWF-1 filters.

Regarding the rank-1 MWF variants without speech covariance matrix reconstruction, the distortionless r1MWF-0 works best on the simulated data while the residual noise power constrained r1MWF-$\mu_{\text G}$ works best on the real data. By changing the trade-off parameter $\mu$ from 0 to \{1,5,10\}, more noise reduction is achieved in the processed signal at the expense of more speech distortion. This results in worse recognition performance in this specific task: WERs increase as $\mu$ increases. Note that for the r1MWF-$\mu_{\text G}$, this trade-off parameter is frequency dependent. In Fig.~\ref{fig-mu}, the spectral gain along frequency and the filtered signals are shown for different parameter values. The r1MWF-$\mu_{\text G}$ has small gain in the low frequencies and puts more weight in the high frequencies, leading to relatively stable level of log-magnitudes as shown in Fig~\ref{fig-mu} (b) and Fig~\ref{fig-mu} (d). The differences in the (time-varying) spectral gain result in different recognition accuracies.

Additional improvement is observed with the speech covariance matrix reconstruction process. On the real test data, the WER is reduced from 8.89\% to 8.09\% for the r1MWF-$\mu_{\text G}$-evd and 7.71\% for the r1MWF-$\mu_{\text G}$-gevd. Overall, the r1MWF-$\mu_{\text G}$-gevd gives the best result. It achieves a 40\% relative WER reduction compared with the baseline WDAS beamformer on the real test set and a 15\% relative WER reduction compared with the GEV-BAN method.

\begin{table}[!ht]
\caption{\pfminor{WERs (\%) achieved by the DNN-sMBR system trained on noisy data from channel 5. These filters are computed from the \emph{correct} masks. The percentages in brackets denote the relative WER changes from the results obtained with the BLSTM predicted masks. The best result for each dataset is in bold.}}
\label{tableOracle}
\begin{center}
\begin{tabular}{|l|l|l|l|l|}
  \hline
  \multirow{2}{*}{Dataset} & \multicolumn{2}{c|}{dev} & \multicolumn{2}{c|}{test} \\ \cline{2-5}
                & ~~~~simu & ~~~~~real & ~~~~simu  & ~~~~~real   \\ \hline \hline
  MVDR     & ~6.25 (-10\%)	&~7.01 (+~2\%)&~7.18 (-17\%) &10.90 (+~6\%) \\ \hline
  GEV-BAN       & ~5.87	(-19\%)&~7.95 (+16\%)&~6.51 (-29\%)	&11.94 (+14\%) \\ \hline
  GEV           & ~6.70	(-11\%)&~8.35 (+18\%)&~7.42	(-26\%)&12.84  (+22\%)\\ \hline
  MWF           & ~7.67	(-31\%)&~8.65 (-~7\%)&~8.36	(-33\%)&15.91  (-~2\%)\\ \hline
  VS          & ~{\bf 4.92} (-~9\%)&~{\bf 6.29} (- 4\%) &~5.83 (-~8\%)&10.43 (+~2\%) \\ \hline \hline

  r1MWF-0               & ~5.00 (-14\%)&~6.44 (-~4\%)& ~{\bf 5.74} (-18\%)& 11.16 (-~2\%)\\ \hline
  r1MWF-1               & ~5.00	(-15\%)&~6.49 (-~3\%)& ~5.75 (-19\%)&11.23 (-~2\%)\\ \hline
  r1MWF-$\mu_{\text G}$     & ~5.81	(-10\%)&~6.71 (+~4\%)&~6.68 (-17\%)&10.94 (+~6\%) \\ \hline \hline
  r1MWF-$\mu_{\text G}$-evd & ~5.65	(-~7\%)&~6.32 (+~3\%)&~6.38 (-16\%)&10.03 (+~8\%)\\ \hline
  r1MWF-$\mu_{\text G}$-gevd &~5.56	(-~7\%)&~6.36 (+~5\%)&~6.24 (-~9\%)&{\bf 10.00} (+14\%) \\ \hline
\end{tabular}
\end{center}
\end{table}

\pfminor{An interesting experiment is to check the performance of these filters using the \emph{correct} masks instead of the predicted ones. This presumably would help to partially discriminate the error rate caused by covariance estimation errors and limitations of the multichannel linear filters themselves. The correct masks for the simulated data are well defined, however, the ground truth underlying the real data is not readily available. The method in~\cite{2016chime4} is adopted for the ground truth estimation for real data and then the masks are calculated using (\ref{eq:ibmx}) and (\ref{eq:ibmn}). The recognition results are summarized in Table~\ref{tableOracle}, with the percentages in brackets denoting the relative WER changes from the results in the left half of Table~\ref{table2}, that are obtained with the BLSTM predicted masks.}

\pfminor{The relative performance between the linear filters is generally consistent with the previous results, though a reduction of WERs on the simulated data is observed and an overall increase of WERs is observed on the real data. For instance, the WERs of GEV-BAN on the test set decrease by 29\% relative on simulated data and increase by 14\% relative on real data. This indicates that GEV-BAN would benefit from better estimated masks on simulated data. This also indicates that the ground truth estimation process is not perfect and GEV-BAN is prone to covariance estimation errors. In comparison, the VS filter is more robust to mask misestimation and achieves the lowest WERs on the development set. Comparing r1MWF-$\mu_{\text G}$-evd and r1MWF-$\mu_{\text G}$-gevd to r1MWF-$\mu_{\text G}$, the rank-1 constraint on the speech covariance matrix still leads to lower error rates. On the real test data, r1MWF-$\mu_{\text G}$-gevd achieves a 16\% relative WER reduction compared with the GEV-BAN method in this case.}

\subsection{Recognition results - Acoustic model trained on enhanced data}

\begin{table}[!ht]
\caption{WERs (\%) achieved by the DNN-sMBR system trained on enhanced data. The best result for each dataset is in bold.}
\label{table3}
\begin{center}
\begin{tabular}{|l|c|c|c|c|}
  \hline
  \multirow{2}{*}{Dataset} & \multicolumn{2}{c|}{dev} & \multicolumn{2}{c|}{test} \\ \cline{2-5}
                & simu & real       & simu  & real   \\ \hline \hline
  Noisy         & 11.43& 12.53  & 14.15 &23.52 \\ \hline
  MVDR          & ~6.80	&~6.97	&~8.61	&11.58  \\ \hline
  GEV-BAN       & ~6.59	&~7.14	&~7.43	&10.62  \\ \hline
  GEV           & ~6.83	&~7.01	&~7.70	&~9.91  \\ \hline
  VS          &\bf ~5.59 &~6.55  &~6.30	&11.12    \\ \hline \hline

  r1MWF-0       & ~5.95	&~6.83	&~6.87	&12.55    \\ \hline
  r1MWF-1       & ~6.72	&~7.45	&~7.70	&13.74  \\ \hline
  r1MWF-$\mu_{\text G}$     & ~6.89	&~7.35	&~7.82	&12.07    \\ \hline \hline

  r1MWF-1-evd   & ~5.92	&~6.66	&~6.96	&12.32    \\ \hline
  r1MWF-1-gevd  & ~5.65	&~6.48  &\bf ~6.13 &11.19    \\ \hline
  r1MWF-$\mu_{\text G}$-evd & ~5.76	&~6.13	&~7.26	&10.33   \\ \hline
  r1MWF-$\mu_{\text G}$-gevd &~5.79	&\bf ~6.04 &~6.48 &\bf ~9.52   \\ \hline
\end{tabular}
\end{center}
\end{table}

In the second experiment, the acoustic model is retrained with the filtered training data. The WERs are shown in Table~\ref{table3}. They are comparable to the left half of Table~\ref{table2} in the sense that the amount of training data is the same.

On the real data, all linear filters generally achieve higher error rates than in the first experiment, except for the GEV filter. On the simulated data, the WERs are generally lower. The proposed r1MWF-$\mu_{\text G}$-gevd is still the best on real data. Note that retraining the acoustic model every time is rather time-consuming and not efficient in practice. The results here provide a strong argument for noisy training, that extends the argument made specifically for the GEV-BAN in~\cite{2016icasspBLSTM}.

\subsection{Analysis}

The above results suggest that neither speech distortion nor noise reduction is straightforwardly correlated with the speech recognition performance. Indeed, the GEV introduces more speech distortion than the theoretically distortionless MVDR but it performs better in the second experiment. The r1MWF-5/10 are supposed to deliver more noise reduction than the r1MWF-0 but they give higher WERs.

\begin{figure}[!t]
	\centering
	\includegraphics[width=0.82\linewidth]{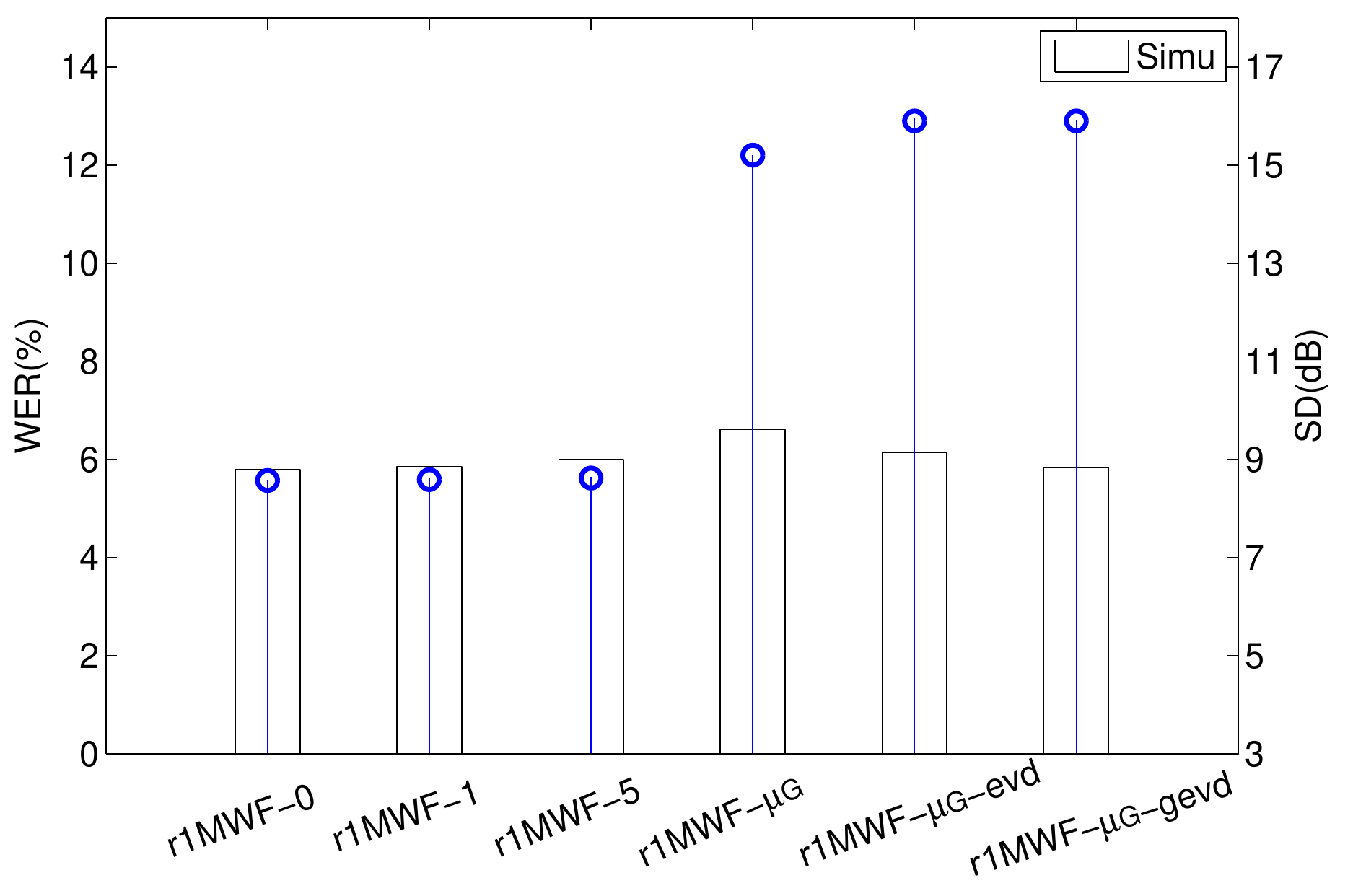}
	\caption{WERs achieved on the acoustic model trained on utterances from all 6 channels and SD scores of the r1MWF variants. WERs are represented by white bars and SD scores are marked by circles.}
	\label{fig4}
\end{figure}

In the following, we investigate the rank-1 MWF variants and their WERs achieved on the noisy acoustic model trained on utterances from all 6 channels. In Fig.~\ref{fig4}, the relation between the WERs and the speech distortion scores is shown. The frequency-weighted log-spectral Signal Distortion (SD) metric~\cite{2011SDevaluation} is defined as
\begin{equation}\label{eq:sdmetric}
\text{SD}=\frac{1}{L}\sum_{l=1}^{l=L}\sqrt{\sum_{k=1}^{k=K}\text{ERB}(k)\left(10\log_{10}\frac{\phi_o}{\phi_i}\right)^2\pfmarker{\text{d}k}}
\end{equation}
where $L$ is the number of frames, $\phi_o$ and $\phi_i$ are respectively the processed speech power spectrum and the clean speech power spectrum, and $\text{ERB}(k)$ is the frequency-weighting factor giving equal weight to each auditory critical band. The SD scores are computed and averaged on the simulated test data. We observe that the r1MWF-$\mu_{\text G}$ introduces much larger distortion than the r1MWF-0/1, from about 9 dB to 16 dB. But the WER only increases slightly. Clearly, there is \pfminor{no strong} correlation between the two.

\begin{figure}[!t]
	\centering
	\includegraphics[width=0.82\linewidth]{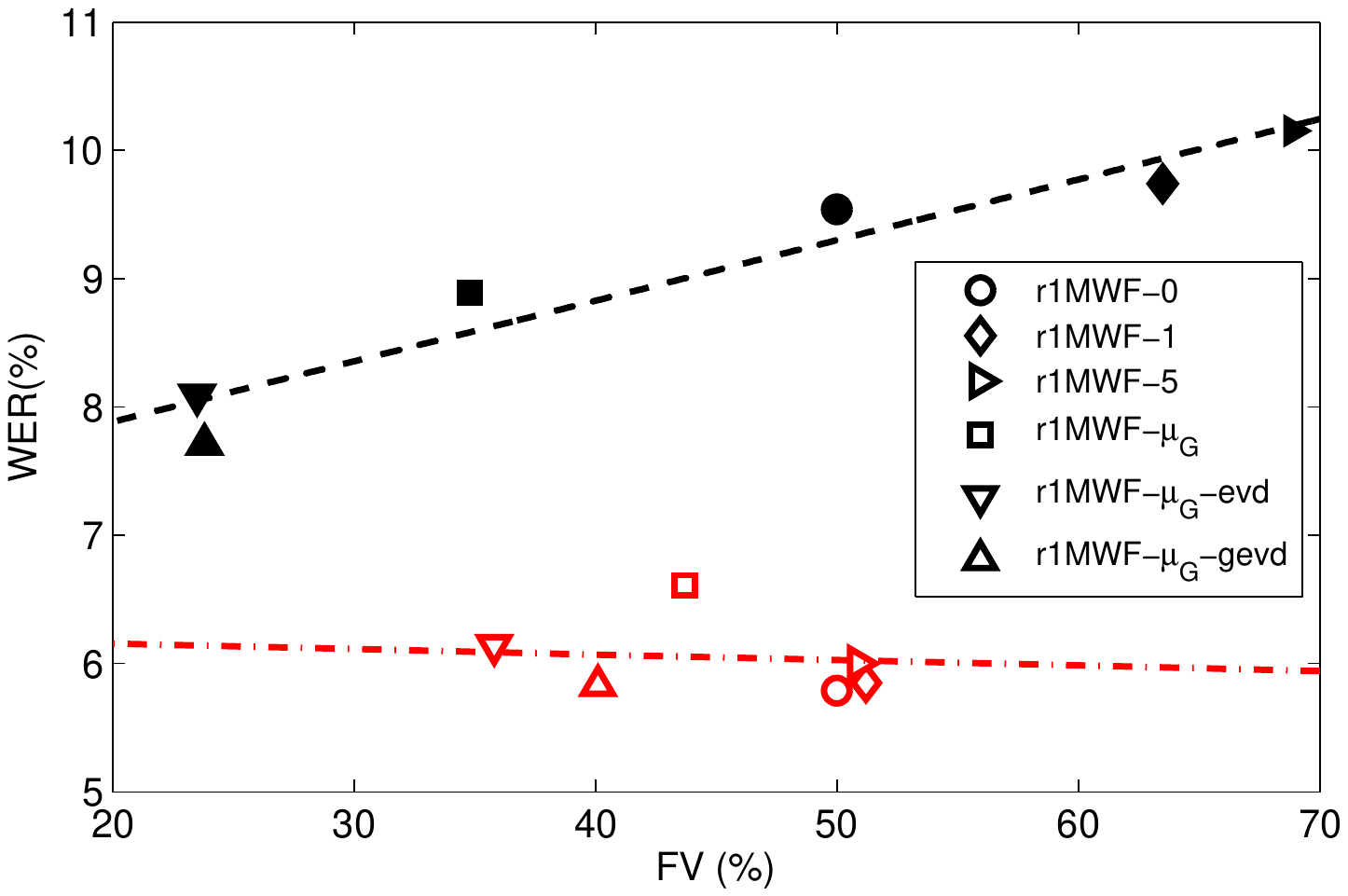}
	\caption{ \pfmarker{Relation between FV and WER for the real (solid black markers) and simulated (hollow red markers) data. The dashed lines show the linear regression results separately. A line with positive slope means positive correlation.}}
	\label{fig5}
\end{figure}

In order to explain the recognition performance, we investigate the variance of the input features corresponding to each HMM state in Fig.~\ref{fig5}. The intuition is that smaller \pfminor{Feature Variance (FV)} implies an easier classification task for the neural network acoustic model. We expect the constant residual noise power property of the r1MWF-$\mu_{\text G}$ to translate into a smaller FV for the processed speech. The HMM state corresponding to each feature vector is first obtained by forced alignment on enhanced data separately. Note that the alignments of the simulated data can be obtained using the clean speech, nevertheless, similar results are observed here. The FV is calculated over all the feature vectors belonging to each HMM state for each method
\begin{equation}\label{eq:fv}
  V(j)=\frac{1}{I}\sum_{i=1}^{I}\text{Var}(i,j)
\end{equation}
where $\text{Var}(i,j)$ means the variance of the $i$th feature in the $j$th state. We pick the FV of the r1MWF-0 as a baseline and define the metric
\begin{equation}\label{eq:featvar}
  \text{FV}=\frac{100}{\sum_{j}c_j}\sum_{j=1}^{J}{\pfmarker{c_j\cdot\mathbb{I}}(V_{\text{test}}(j)>V_{\text{baseline}}(j))}
\end{equation}
that is the \pfmarker{weighted} percentage of states for which the FV is larger than the baseline. \pfmarker{$c_j$ denotes the number of occurrences of the $j$th state. $\mathbb{I}(\cdot)$ is an indicator function the value of which is 1 for true arguments and 0 for false.} For a comparable method, the value is expected to be around 50\%. On the real data, the r1MWF-1 and r1MWF-5 have higher percentages (\pfmarker{62.7\% and 68.4\%}) and corresponding higher WERs. For the r1MWF-$\mu_{\text G}$-evd and r1MWF-$\mu_{\text G}$-gevd, lower percentages (\pfmarker{23.1\% and 23.6\%}) correlate with lower WERs. However, the correlation is not always valid on the simulated data as shown by the \pfmarker{r1MWF-$\mu_{\text G}$: it has 43.7\% states} with smaller FV and yet a higher WER than the baseline.

\pfmarker{The FV metric provides another view from the feature side to explain the performance of the constant residual noise filter. Note that a global scale factor only results in a shift in the 0th MFCC value and will not affect the feature variance. The computation of FV also avoids the time-consuming decoding procedure that is required for WER.}

\section{Conclusion}

Multichannel linear filters are generally designed to improve the speech perceptual quality but not specifically to improve the speech recognition accuracy. As a matter of fact, the choice of the optimal filter may be different for different tasks. In the scenario of a single target source, the popular SDW-MWF can be formulated as the rank-1 MWF. We derived a family of rank-1 MWF variants and evaluated their performance for speech recognition in multiple noisy environments. We defined a constant residual noise power constraint to find the trade-off parameter which links the rank-1 MWF filter and the GEV beamformer. We showed that this constraint brings more speech distortion, however, it benefits the speech recognition performance on the real data. To fulfill the underlying rank-1 assumption, speech covariance matrix reconstruction is proposed. The reconstruction based on eigenvectors or generalized eigenvectors subsequently improves the recognition accuracy. With experiments conducted on the CHiME-4 dataset, the final r1MWF-$\mu_{\text G}$-gevd filter achieved a 40\% relative WER reduction compared with the baseline WDAS beamformer on the real test set and a 15\% relative WER reduction compared with the GEV-BAN method. \pfmarker{For future research, we would like to see how the performance is impacted for corpora with higher reverberation time where the narrowband approximation becomes more erroneous.}

In the speech recognition task, it is observed that multi-condition noisy training works well and sometimes outperforms retraining with enhanced data. So when new signal processing methods are applied, a reasonable practice is to process only the test data. Another finding is that the speech perceptual quality is not straightforwardly related to the speech recognition performance. An investigation from the perspective of feature variance is provided. The work puts forward the need for novel signal or feature metrics that correlate better with the WER.

\section{Acknowledgements}

\pfminor{We would like to thank the anonymous reviewers for their constructive comments.} This work was supported by the China Scholarship Council (No. 201604910623). Experiments presented in this paper were carried out using the Grid'5000 testbed, supported by a scientific interest group hosted by Inria and including CNRS, RENATER and several Universities as well as other organizations (see \url{https://www.grid5000.fr}).

\section*{References}

\bibliography{reference}

\end{document}